# Predicting the flow stress and dominant yielding mechanisms: analytical models based on discrete dislocation plasticity

Jianqiao Hu [a, b], Hengxu Song [d, *], Zhanli Liu [c], Zhuo Zhuang [c], Xiaoming Liu[a, b, *], Stefan Sandfeld [d]

[a] State Key Laboratory of Nonlinear Mechanics, Institute of Mechanics, Chinese Academy of Sciences, Beijing 100190, P R China

[b] School of Engineering Science, University of Chinese Academy of Sciences, Beijing 100049, P R China

[c] Applied Mechanics Lab., School of Aerospace Engineering, Tsinghua University, Beijing 100084, P R China

[d] Chair of Micromechanical Materials Modelling, Institute of Mechanics and Fluid Dynamics, TU Bergakademie Freiberg, Freiberg 09599, Germany

*Corresponding Authors: Hengxu.Song@imfd.tu-freiberg.de, xiaomingliu@imech.ac.cn*

**Abstract:**

Dislocations are the carriers of plasticity in crystalline materials. Their collective interaction behavior is dependent on the strain rate and sample size. In small specimens, details of the nucleation process are of particular importance. In the present work, discrete dislocation dynamics (DDD) simulations are performed to investigate the dominant yielding mechanisms in single crystalline copper pillars with diameters ranging from 100 to 800 nm. Based on our simulations with different strain rates and sample size, we observe a transition of the relevant nucleation mechanism from "dislocation multiplication" to "surface nucleation". Two physics-based analytical models are established to quantitatively predict this transition, showing a good agreement for different strain rates with our DDD simulation data and with available experimental data. Therefore, the proposed analytical models help to understand the interplay between different physical parameters and nucleation mechanisms and are well suitable to estimate the material strength for different material properties and under given loading conditions.

1. Introduction

A fundamental understanding of small-scale plastic behavior is of great significance for many micro-scale specimens and devices relevant for engineering applications. Such samples exhibit size-dependent mechanical behavior when their size reaches an internal microstructural length scale. For many metallic devices, the length scale ranges typically from below hundred nanometers up to tens of micrometers. For plasticity in micrometer-size specimens, modelling such deformation becomes amendable to continuum approaches [1-6] that consider the dislocation microstructure in an averaged sense by densities. For polycrystals the crystal plasticity finite element method was used to study strain rate effects on aluminum single crystals [7]. Dislocation-density based models have been used to study strain rate-dependent deformation, where very distinct strain-rate regimes were observed in the stress-strain response [8]. However, for smaller crystals, the discreteness of dislocation starts to play a more and more important role [9, 10]. For example, the probability of dislocations interacting with free surfaces increases, which strongly change the deformation mechanisms [11]. This affects also the dominant nucleation process when the sample dimensions decrease: in the case of low strain rates, both TEM observations [12, 13] and discrete dislocation dynamics (DDD) [14, 15] revealed that for large micrometer-sized samples dislocations are mainly nucleated through Frank-Read source mechanisms, while for samples on the sub-micrometer scale nucleation happens rather by so-called "single-arm sources" (SAS), i.e. a dislocation segment which is pinned inside the sample that terminates on the sample surface.

Such an increased activation of SAS's [16-18] together with an increasing surface area then explains the size-dependent yield strength in sub-micrometer face-centered and body-centered cubic crystals: When the sample size decreases down to the nanometer scale (e.g., ~160nm in [19]), experiments show that dislocations tend to mainly escape through the free surface due to limited space and high image stress. As a consequence, the reduced number of dislocations makes it difficult to form jammed and stable dislocation microstructures during loading. Subsequently, when the sample enters a state of dislocation starvation [20] and the stress level increases, dislocations will be created through "surface nucleation" (SN). DDD simulations have shown [21] that this is the main mechanism for plastic deformation in small pillars with diameter less than 200nm. A number of pioneering works for studying the process SN with DDD and MD simulations has been carried out in [22-26]. Furthermore, as a general rule it was found that the critical stress to trigger a SN lies between the stress of nucleation in the bulk [27] and the stress required to activate a SAS [28]. By performing uniaxial compression tests and nanoindentation experiments [29], dislocations can well be analyzed at the free surface, and it is found that the critical stress to nucleate a full dislocation loop in the bulk is $\sim\mu/8$, while the stresses required to nucleate a half loop at a surface is $\sim\mu/26$ with $\mu$ the shear modulus.

Besides the size effect, the second important influence on details of the plastic deformation is the strain rate. E.g., in [30, 31] it was shown that the yield strength increases with an increasing loading rate as the high strain rate suppresses dislocation pattern formation [32]. 2D DDD simulations [33] have recently shown that increasing the strain/stress rate results in transition from nucleation dominated to drag dominated with dislocation avalanches following a power-law statistics. Experimental studies of copper pillar compression [34] revealed that the strain rate, similar to the sample size, has an apparent effect on the competition between SAS and SN. And also a theoretical study of the activation volume based on the experimentally measured stress demonstrates that SN is predominant in smaller samples or under high stress rate loading.

This competition between size effects and strain rate effects naturally leads to a complex crossover between SAS and SN mechanism. Such interdependencies are very challenging when it comes to material design or reliability in engineering applications where strongly varying sample size are in use and strain rates during service varies. DDD simulations that incorporate SN mechanisms as a local rule [35] can reveal the transition of mechanism under different conditions. However, DDD simulations are computationally very expensive. Therefore analytical models that can predict the sample flow stress and associated mechanism are highly desirable. In this paper, based on DDD simulations of different sample sizes and strain rates, we propose and extend analytical models that can describe SAS and SN. The two analytical models are then combined to predict the transition of yielding due to the different nucleation mechanism. Parameterized by DDD simulations, our analytical model successfully matches the experimental results for different sample sizes and external loading strain rate and thereby helps to understand the complex interaction between all involved model and material parameter.

The paper is structured as follows: our DDD approach is briefly described in Section 2, followed by presenting the simulation results to illustrate the effects of sample size and strain rate in Section 3. In Section 4, the analytical models based on the SN and SAS mechanisms are established to evaluate the flow stress in submicron copper pillars. The results predicted by the two analytical models are then analyzed to clarify the critical size, at which the transition from dislocation multiplication dominated plasticity to the surface nucleation dominated process occurs for quasi-static loading. Furthermore, the two analytical models are also used to predict the dislocation mechanism transition under high strain rate loading. Finally, we summarize and discuss the results in Section 5.

## 2. Simulation procedures

DDD has been widely used to study the dislocation controlled plasticity at submicron scales [36-39]. The DDD method adopted here has been described in detail in our previous work [40, 41], and in what follows we only give the most important information. The general setup of discrete dislocation dynamics is: for each time step (i) apply a load increment on the sample in a strain-rate-controlled manner; (ii) calculate the Peach-Koehler forces acting on all dislocations due to the applied load and the mutual interaction, (iii) compute nodal velocities based on the Peach-Koehler forces, (iv) compute the updated segment positions by time integrating the equation of dislocation motion, and (v) apply so-called local rules which govern non-elastic effects such as intersections and locks. In our simulations, we consider dislocations escaping from the free surfaces as well as surface effects on the stress fields (image forces). Specifically, image forces are considered through Lothe's analytical solution [42, 43], an approach that has been often used [44, 45] and does not require a computationally expensive finite element calculation. The plastic strain rate of the simulated crystal is calculated from the dislocation velocity and slip system geometry (i.e. the Burgers vector and the normal vector of slip surface) [46, 47], and for the here considered compressive tests the external stress is calculated through $\sigma = E(\varepsilon - \varepsilon^p)$. There, the engineering stress $\sigma$ is the compressive axial stress, and the engineering strain $\varepsilon$ is the compressive axial strain. Subsequently, we will consider various loading strain rates in the DDD simulation. Wang et al. [48] investigated if inertial effects need to be considered in such DDD simulations and concluded that these become relevant when the strain rate is higher than $10^3$ s$^{-1}$. In Wang's study, it has also been demonstrated that the relativistic effects (i.e., the change in effective mass when the dislocation velocity approaches the speed of sound) play a minor role since very few dislocations move at a speed greater than 25% of the speed of sound. Therefore, the present DDD model includes the full dynamical equation of dislocation motion without relativistic effects, $m\dot{v} + Bv = f$, where $m = \rho_m b^2$ is the effective mass of dislocations per unit length with $\rho_m$ the mass density, $b$ the magnitude of the Burgers vector, and $B$ is the viscous drag coefficient of dislocation motion. $v$ is the dislocation velocity and its time derivative $\dot{v}$ is the acceleration. Atomistic simulations [25, 26] have shown that surface nucleation is the dominant mechanism responsible for yielding for small volumes. Based on these simulation results, a surface nucleation model was proposed [21, 35] which has also been adopted in DDD simulations: the model states that within a time span $\Delta t$, the probability of a surface nucleation event is given by

$$P = \nu_0 \exp\left[-\frac{Q(\sigma,T)}{k_B T}\right] \times \left(\frac{S}{b^2}\right) \times \Delta t . \tag{1}$$

Here, $\nu_0$ is the attempt frequency and $k_B T$ is the thermal energy with $k_B$ the Boltzmann constant. The activation energy $Q(\sigma,T)$ for SN depends on stress and temperature and takes the form [25]

$$Q(\sigma,T) = \left(1 - \frac{T}{T_m}\right) \cdot Q_0(\sigma) \text{ with } Q_0(\sigma) = A \cdot \left(1 - \sigma/\sigma_{\text{athm}}\right)^{\alpha_0}, \tag{2}$$

where the constants and parameters $T_m$, $A$, $\alpha_0$, and $\sigma_{athm}$ are given in the appendix. $\sigma = |\sigma_{zz}|$

is used in the DDD simulation of uniaxial compression. $\frac{S}{b^2}$ gives the potential number of SN sites with $S$ being the surface area of the sample. The probability of an SN event is estimated at each time increment. Then, for a time span for which the nucleation condition is satisfied, an arc-shaped dislocation loop will nucleate at a random site of the free surface. In the DDD model, nucleation is only allowed to occur on the 1/2<110>/{111} type slip systems. In addition, the initial radius of the newly nucleated arc-shaped dislocations is chosen as 30$b$.

In our DDD simulations, the side length $D$ of the copper pillars varies from 100 nm to 800 nm with a quadratic cross section. The aspect ratio $H/D$ of all pillars is fixed to 2 with $H$ being the pillar height. Prior to loading, a relaxation process was performed, which can be interpreted as a thermal annealing process [21, 43, 49, 50]. In this process, straight dislocation lines as well as dislocation loops are randomly generated on different slip systems. Then the dislocation structure evolves without external loading until equilibrium is reached. After this initial relaxation procedure, we obtained samples with an initial dislocation density consistent with the experimental values ($10^{13}$~$10^{15}$ m$^{-2}$) in FIB fabricated micropillars. All pillars are loaded at the top along the $(00\bar{1})$ direction and the lateral surfaces are traction free. The imposed strain rate ranges from $10^4$ s$^{-1}$ to $10^8$ s$^{-1}$. Note that DDD simulations are typically done at much higher strain rates than experimental ones. Due to this offset, loading in our simulations is not considered as shock loading and it is not necessary to use dislocation elastodynamics [51, 52]. Considering the randomness of the initial microstructure, for pillars with the same size in DDD calculations, more than three realizations of different initial dislocation configurations are simulated to reveal the stochastic scatter. The material properties of copper used in the simulation are listed in Table A1 in Appendix A. To ensure the convergence of numerical results, the time step d$t$ used in the DDD simulation is $10^{-12}$ s. In the DDD simulations, the stress strain curves strongly fluctuate in the flow stage, therefore we used a commonly used method to define the flow stress: take the stress value of the last simulation increment, find at least two smaller strains that correspond to the same stress value, i.e., find the periodicity of the fluctuation. The average of all stress values within the periodicity is defined as the flow stress.

3. **DDD simulation results**

The mechanical responses of pillars of different sizes ranging from 100nm to 800nm under a specific strain rate ($10^6$ s$^{-1}$) are shown in Figure 1(a). The stress increases elastically at the initial stage, and ends with two distinctly different stress levels, depending on the sample size. In large pillars (*D*=800 nm), soon after the initial elastic increase, yielding takes place at a relatively low stress level of ~500MPa. By contrast, the stress reaches values as high as ~1700MPa in the small pillars (*D*=100 nm) after an extensive elastic period. For pillars with diameter 200 nm and 400 nm, the stress-strain relation exhibits the above two behaviors across sample realizations. Figure 1(b) shows typical dislocation structures in pillars with different sizes. There it can be seen that in large pillars dislocations can initially form jammed configuration, and these complex networks will continuously evolve until stable pinning points are formed inside. Consequently, the operation of internal single-arm dislocation sources generates enough plasticity to accommodate the external loading, featuring low stress levels. However, in small pillars, pre-existing dislocations are quite unstable since it is difficult to form stable pining points in a limited space, such that these dislocations escape from the free surfaces of pillar featuring the dislocation starvation mechanism observed in the experimental studies [20, 53]. Subsequently, new dislocations get nucleated from free surfaces as the loading continues.

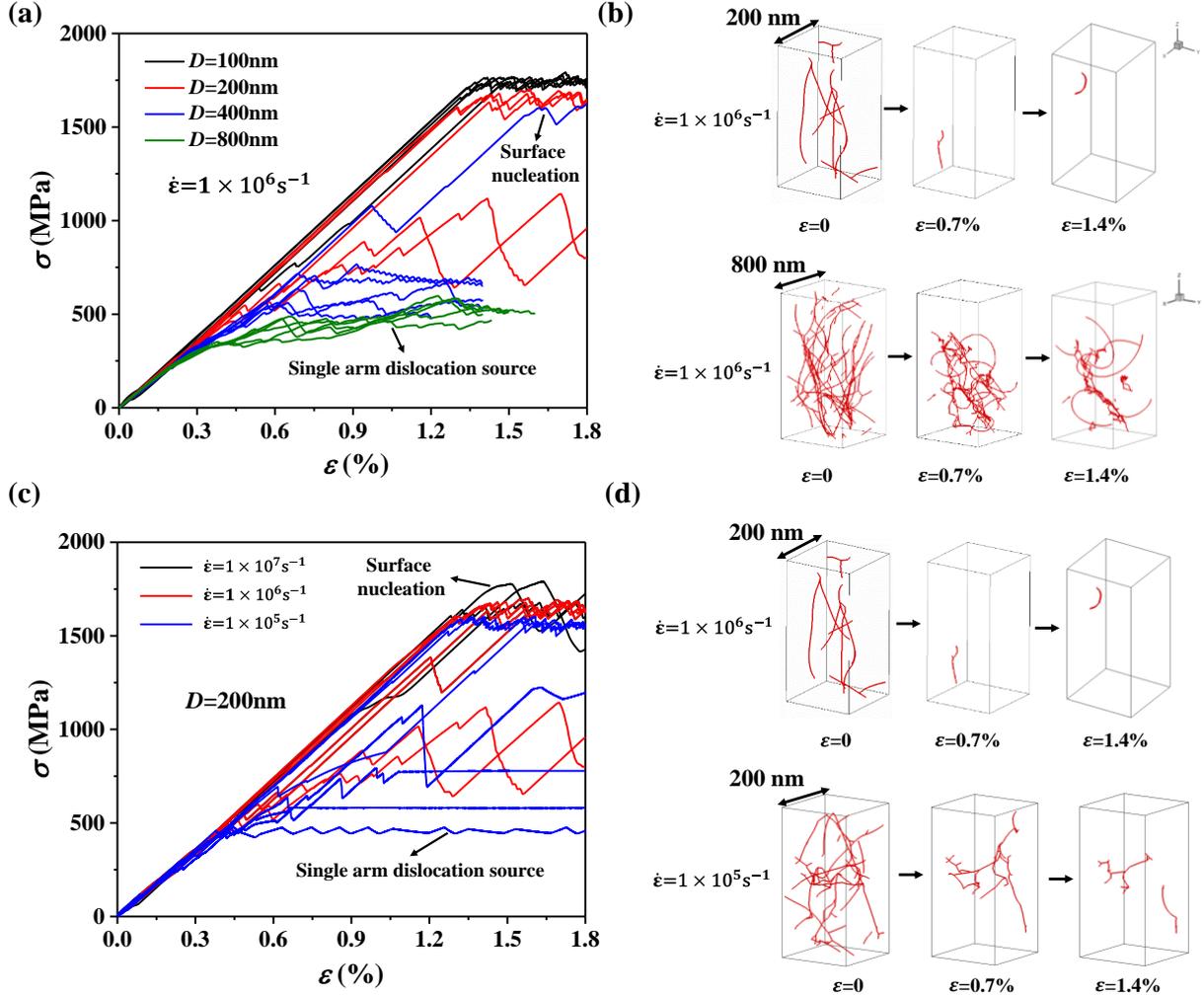

Figure 1 (a) Stress-strain curves and (b) evolution of dislocation structures in pillars with different sizes at an applied strain rate of $10^6$ s$^{-1}$; (c) stress-strain curves and (d) evolution of dislocation structures in pillars with $D$=200nm for various strain rates.

For the pillar of size 200nm, simulations of different strain rates were carried out. It can be seen in Figure 1(c) that the larger strain rate results in higher flow stress. Typical evolutions of dislocation microstructures at different rates are shown in Figure 1(d): when the strain rate is high, stable SAS cannot be formed, therefore, SN will occur subsequently. By contrast, for the smaller strain rate, stable SAS exists and accommodate the plastic flow. Through the above analysis, it can be seen that either decreasing the sample size or increasing the strain rate would result in a change of the yielding mechanism from SAS to SN. A complete investigation of the influence of all realistic sample sizes and loading rates on the dominant dislocation mechanism would require a huge number of DDD simulations and we therefore introduce a phenomenological, mechanism-based model that is well able to represent average features of the investigated pillar systems.

## 4. Theoretical analysis

We now introduce two analytical models – for SN and for nucleation by SASs – to predict the flow stress in the two different nucleation regimes. By comparing and combining the two models we then analyze the critical size, the strain rate and dislocation density dependency.

### 4.1 Surface nucleation dominated model

From the stress-strain curves and the evolution of corresponding dislocation structures shown in Figure 1, it can be seen that the SN dominates the plastic behavior in pillars with small size or under high strain rate. The flow stress can be evaluated by the activation stress at the onset of SN processes since single SN event can trigger yielding in small pillars. The SN dominated flow stress can then analytically be evaluated based on our probabilistic criterion of dislocation nucleation. During the time span $t_{life}$, i.e., the time from dislocation nucleation at the free surface until the dislocation escapes the volume, the number of dislocation nucleation events $P_s$ can be calculated through Eq. (1) where $\Delta t$ is replaced by $t_{life}$. The plastic strain increment generated by dislocation motion during the time span $t_{life}$ is

$$\Delta \gamma = \frac{\beta \cdot b \cdot S_c}{V}, \tag{3}$$

Where $S_c = D^2/\cos\theta$ is the slip plane area, and $\theta$ is the angle between the slip plane normal and the loading direction. $\beta \cdot S_c$ represents the estimate of the real slipped area, where we assume $\beta \approx 0.5$ (however, the resulting flow stress is mostly insensitive to variations of $\beta$ in the range from 0.3 to 1.0). $M$ is the Schmid factor which is 0.4082 for the allowed nucleation on the $1/2<110>/\{111\}$ type slip systems; $V$ is the sample volume. Considering that strain hardening is virtually absent in micropillars [14], we assume that dislocations originating from SN can generate enough plasticity to accommodate the deformation induced by the increase of the external loading, therefore the time span $t_{life}$ can be calculated as

$$t_{life} = \frac{\Delta \varepsilon^p}{\dot{\varepsilon}},  \qquad (4)$$

where $\Delta \varepsilon^p$ is the resulting axial plastic strain increment given by $\Delta \varepsilon^p = \Delta \gamma \cdot M$, $\dot{\varepsilon}$ is the axial loading strain rate. $t_{life}$ should ensure the uninterrupted dislocation nucleation from free surfaces, namely, once a SN event takes place (i.e., a half dislocation loop is introduced into the sample) and the half loop exits the sample again, the next SN emerges subsequently. Thus, during the time span $t_{life}$, the probability that a nucleation event takes place is 1.

Based on the above analysis, the flow stress for SN dominated systems can be solved implicitly. Firstly, given the initial parameters required in the probabilistic model in Eq. (1) and the loading rate $\dot{\varepsilon}$, the time span and the nucleation event can be estimated for an initial stress value. Then, the axial stress gradually increases by a small amount until a surface nucleation event is triggered (i.e., $P_s \geq 1$). Finally, the flow stress can be obtained. The related details (flow chart, simulation parameters) can be found in Appendix A. The used Python code for the continuum simulations is freely available as supplementary material. In Figure 2, the flow stress of pillars with diameter 100nm at different strain rates calculated by both the analytical model and DDD simulations are presented. Without any additional fitting parameters, the analytical model predicts the flow stress for pillars under different strain rate loading well within a relative error of less than 7%.

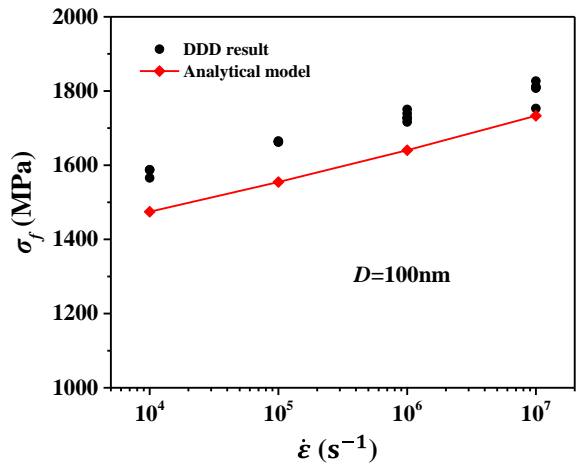

Figure 2 Comparison of simulation results and analytical solutions in pillars with diameter 100nm under different strain rates.

## 4.2 Single-arm source dominated model

For large pillars ($D>400$nm) and low imposed external strain rate, experiment and simulation both showed that single-arm dislocation sources control the plastic flow in submicron single crystals [13, 54-56]. A formulation for the critical resolved shear stress (CRSS) in single crystal pillars under quasi-static loading has been already investigated by a theoretical model [28, 41]. This model has been further extended by Hu et al. [35] towards different strain rates, as

$$\tau = \underbrace{\tau_0 + \alpha\mu b\sqrt{\rho} + \frac{k\mu b}{\lambda_s}}_{\tau_{static}} + \frac{2B\dot{\varepsilon}V}{Mb^2\lambda_c \cdot n}, \qquad (5)$$

where $\tau_{static}$ represents the conventional SAS controlled CRSS under quasi-static loading and consists of the lattice friction stress $\tau_0$, a so-called Taylor term which governs the interaction of dislocations with density $\rho$, and a line tension contribution where $\lambda_s$ is the average effective straight source length. While the Taylor term is most commonly used in the macroscopic models it has also been frequently used in sub-micron plasticity [57] where it is often connected to the mean free path of a dislocation. $\alpha$ and $k$ are constants and taken as 0.5 and 1.0 as in [18, 28]. The last term governs strain rate effects where $\lambda_c$ is the effective curved source length which can be estimated as the sample size $D$ [58]. $n$ is the number of stable SASs and can be estimated by the pillar size and the dislocation density

$$n = \text{Integer}\left[\frac{\pi\rho DH}{4}\right]. \qquad (6)$$

The evolution of dislocation density at the sub-micrometer scales is represented as [41],

$$\frac{d\rho}{d\gamma} = \frac{1}{2b\lambda_s} - \frac{2\cos^2(\theta/2)}{bD} + \frac{k_f\sqrt{\rho}}{b} - \frac{y_c}{b}\rho, \qquad (7)$$

where $\lambda_s$ can be calculated by the model original from [28], $k_f$ and $y_c$ are dimensionless constant and effective mutual annihilation distance, respectively.

For SAS dominated plasticity, the external loading forces the internal dislocation sources to generate sufficient plasticity such that the pillar exhibits a constant plastic flow. The stress level when internal SASs prevail is generally not high enough for the SN to become important. Corresponding to the typical dislocation density ($10^{12}$~$10^{13}$ m$^{-2}$) in pillars with diameter 800nm, the number of SASs given by Eq.(6) is approximately 1~10. As shown in Figure 3(a), the flow stress of pillars with diameter 800nm predicted by Eq. (5) is presented as a function of the number of SASs and strain rates. Firstly, it can be seen with the decrease of the strain rate, for example when $\dot{\varepsilon}=1000$ s$^{-1}$, the flow stress is insensitive to the number of SASs. This is because at a relatively low strain rate loading, the activation of a few dislocation sources is sufficient to maintain continuous plastic flow. While for high strain rate loading, see $\dot{\varepsilon}=10^6$ s$^{-1}$ in Figure 3(a), the flow stress becomes sensitive to the number of SASs, e.g., fewer SASs results in higher flow stress.

For a pillar with a given size, the minimum number of dislocation sources that are required to maintain the SAS dominated plasticity is determined as the following : first, the size dependent CRSS at a certain strain rate is calculated by Eq.(5) through initially setting the number of SAS to 1; then it is checked if the dislocation velocity $v_d$ exceeds the shear wave velocity $C_s$ in the material. If $v_d > C_s$, then more SASs are required, since increasing the dislocation density (i.e., the number of SASs) in the Orowan equation $\dot{\gamma}=\rho b v_d$ can decrease the required dislocation velocity while maintaining a certain plastic deformation rate. More details can be found in Appendix A. The used Python code for the continuum simulations is freely available as supplementary material. In Figure 3(b), the analytical predictions by the SAS dominated model, by the SN model and the DDD simulation results under a certain strain rate (i.e., $\dot{\varepsilon}=10^6$ s$^{-1}$) are compared. The discreteness of the flow stress in DDD simulations is due to the difference in the number of stable SASs (dislocation densities). When the stable dislocation density is high, i.e., there are more stable SASs, the SAS model predicts a lower bound of the yield strength, for example the red curve for $\rho_s=10^{14}$ m$^{-2}$ in Figure 3(b). When the dislocation density decreases to a limit which can only provide the minimum number of stable SASs that is required by the SAS mechanism, our SAS model predicts the so-called upper bound of the yield strength for the SAS mechanism. This is shown in Figure 3(b) through the green line for different pillar size $D$. The fluctuations there are caused by the fact that for the same number of the stable SAS's (compare, e.g., *n* for *D*=400 nm and 500nm), the strain rate term (the last term in Eq. (5)) has a stronger influence on the yield strength for a large sample than for a small sample (note, that when computing the yield strength without considering the strain rate term, the yield strength as a function of the sample size *D*, based on dislocation densities that correspond to *n*, is smooth and shows the typical strengthening size effect). In Figure 3(b), we also show the prediction of the flow stress by the SN mechanism (black line). It can be seen that for our DDD results of *D*>400nm, the yielding mechanism is SAS. For *D*=400 nm and 200nm, the yielding mechanism in some of the simulations (whose strength is between the two bounds) is governed by SASs while it is the SN mechanism for the rest (whose strength is close to the SN model prediction); for *D*=100nm, we also observe the SN mechanism for all DDD simulations since SN usually requires a higher activation stress but here it predicts a smaller yield strength than the SAS mechanism, therefore the SN mechanism is favored. The evaluation of the yielding mechanism based on the two analytical models here is also consistent with the evaluation made in Figure 1(a) based on the observation of the dislocation microstructure.

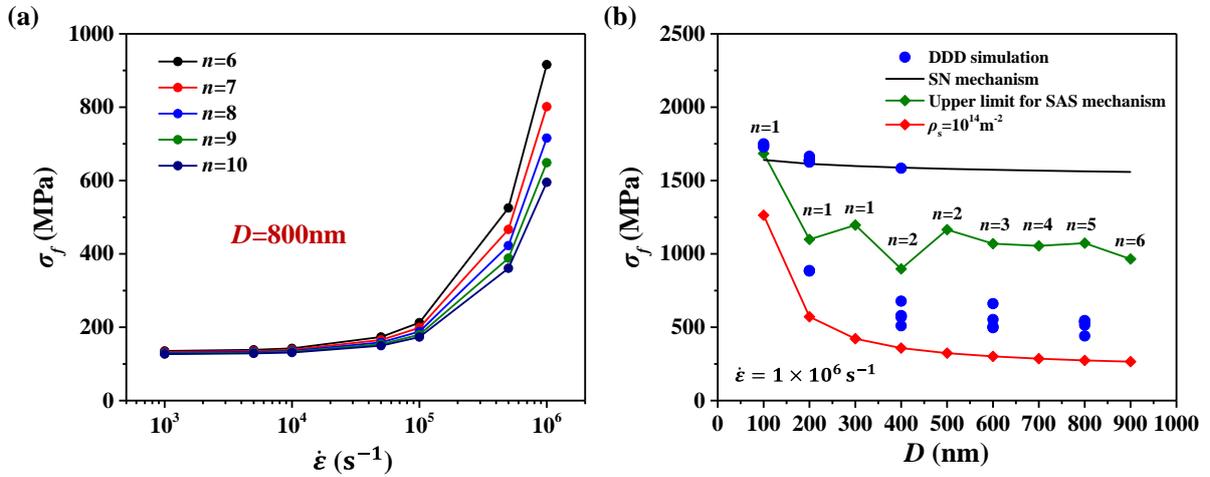

Figure 3 (a) The flow stress as a function of the number of SASs and strain rates for a pillar diameter of *D*=800nm. (b)The flow stress obtained from DDD simulation and from the analytical models.

**4.3 Critical size for the transition of dislocation mechanism under quasi-static loading**

What is the critical size when the transition between the two nucleation mechanisms takes place under quasi-static? We compare the flow stress calculated by the two analytical models at a quasi-static strain rate: the flow stress from the SN model is calculated at the strain rate of 0.001 s$^{-1}$ while the last term in Eq. (5) can be neglected in the SAS model. Initial dislocation density $\rho_0$ is set to an experimentally observed value $10^{14}$ m$^{-2}$ [11] in the SAS model, and the dislocation density evolves according to Eq.(7). In Figure 4, the flow stress calculated by the two models as well as experiment results from [11, 34, 58-60] are shown. The analytical predictions match the experimental results very well. Moreover, an intersection point between the SAS prediction and the SN prediction is observed. There, a critical sample size is seen to be ~110nm, which indicates the transition between the two yielding mechanisms. For sample diameters below ~110nm the flow stress evaluated by the SN model is less than that for the SAS model. Therefore, the activation of a SAS is more difficult than dislocation nucleation from a free surface such that SN is favored. By contrast, for samples larger than this critical size, plasticity tends to be generated by the operation of pre-existing dislocation sources, i.e., the SAS mechanism. Different mechanisms for plastic deformation and the formation of dislocation structures for various sample sizes have been observed in many experiments. Through transmission electron microscopy (TEM), Shan et al. [19] reported that for Ni pillar with diameter 160nm, the pre-existing dislocations progressively left the pillar and then are accompanied by new dislocations (which corresponds to SN); while for pillars with diameter of 290nm, there still exists dislocations (which can roughly be related to SAS's) after a period of compression. The critical size in BCC single crystal Mo is ~200nm [61]. Thus, our predictions agree well with these experimental observations.

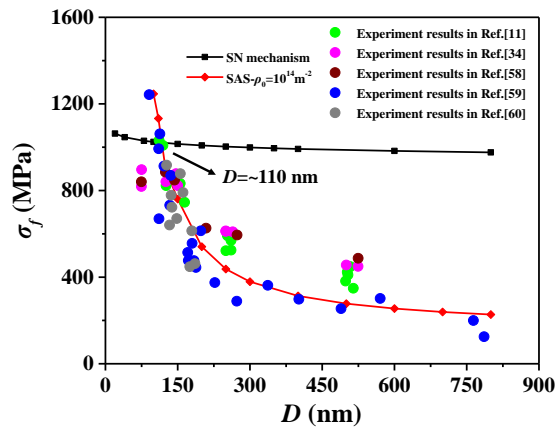

Figure 4 The flow stress of Cu pillars with diameter ranging from 70 nm to 800 nm by the SN and SAS models and relative experimental results.

**4.4 Strain rate effect on the transition of dislocation mechanism**

In this section, the transition from SAS to SN triggered by the increasing strain rate is studied for two different sample sizes. The flow stress is predicted by the analytical models through the following steps: (1) for a certain initial dislocation density, the evolution of dislocation density is calculated by Eq.(7), and the number of SASs can be estimated by Eq.(6); (2) the minimum number of dislocation sources required to maintain SAS dominated plasticity can also be obtained by the same way as shown in Figure 3(b); (3) the relevant yielding mechanism (nucleation by SAS or SN) can be determined by comparing the number of actual and required SASs, the flow stress is then evaluated by the corresponding analytical model. In the analytical model, the initial dislocation densities are set to $10^{13}$ m$^{-2}$ and $10^{14}$ m$^{-2}$ for the two sample sizes, which is consistent with the range of initial dislocation density ($10^{13}$~$10^{14}$ m$^{-2}$) in DDD simulations. Note that in DDD simulations, initial dislocation density stands for the density of the relaxed configuration, not the starting configuration of straight segments before the relaxation. Therefore, in the simulation it is difficult to precisely control the initial density and there is some scatter in the data. In Figure 5(a) for the sample size of 400nm, the flow stress predicted by the SN mechanism is also presented. Both samples exhibit a transition from SAS to SN triggered by the increasing strain rate as shown in Figure 5(a) and (b), and the transition is delayed for higher dislocation density (from red to the blue curve). Despite the success in the consistent trend and good agreement between analytical predictions and DDD results, it is also noticed that the larger dispersion of flow stresses exists in the sample of smaller size, and some results from DDD simulations deviate from the analytical model. To interpret the discrepancy, the evolution of dislocation structures at two strain rates in a pillar of $D$=400nm is illustrated in Figure 5(c), the number of dislocation sources is quite limited in this small-sized pillar, and some dislocations are pinned inside. More importantly, the SASs are activated intermittently, which deviates from the assumption in the analytical model that SASs move continuously in the plastic flow stage. Therefore, under this circumstance, the analytical model prediction of SAS mechanism is the lower bound of the DDD simulation results. In contrast, for pillars of $D$=800nm, as shown in Figure 5(d), the activation of SASs exists and continues in each deformation stage, featuring a relatively more stable deformation in larger pillars. Furthermore, it is also observed that, for the dislocation densities considered here (which are also reasonable values observed in experiments), high strain rates of $\geq 10^8$ s$^{-1}$ will always trigger the SN mechanism.

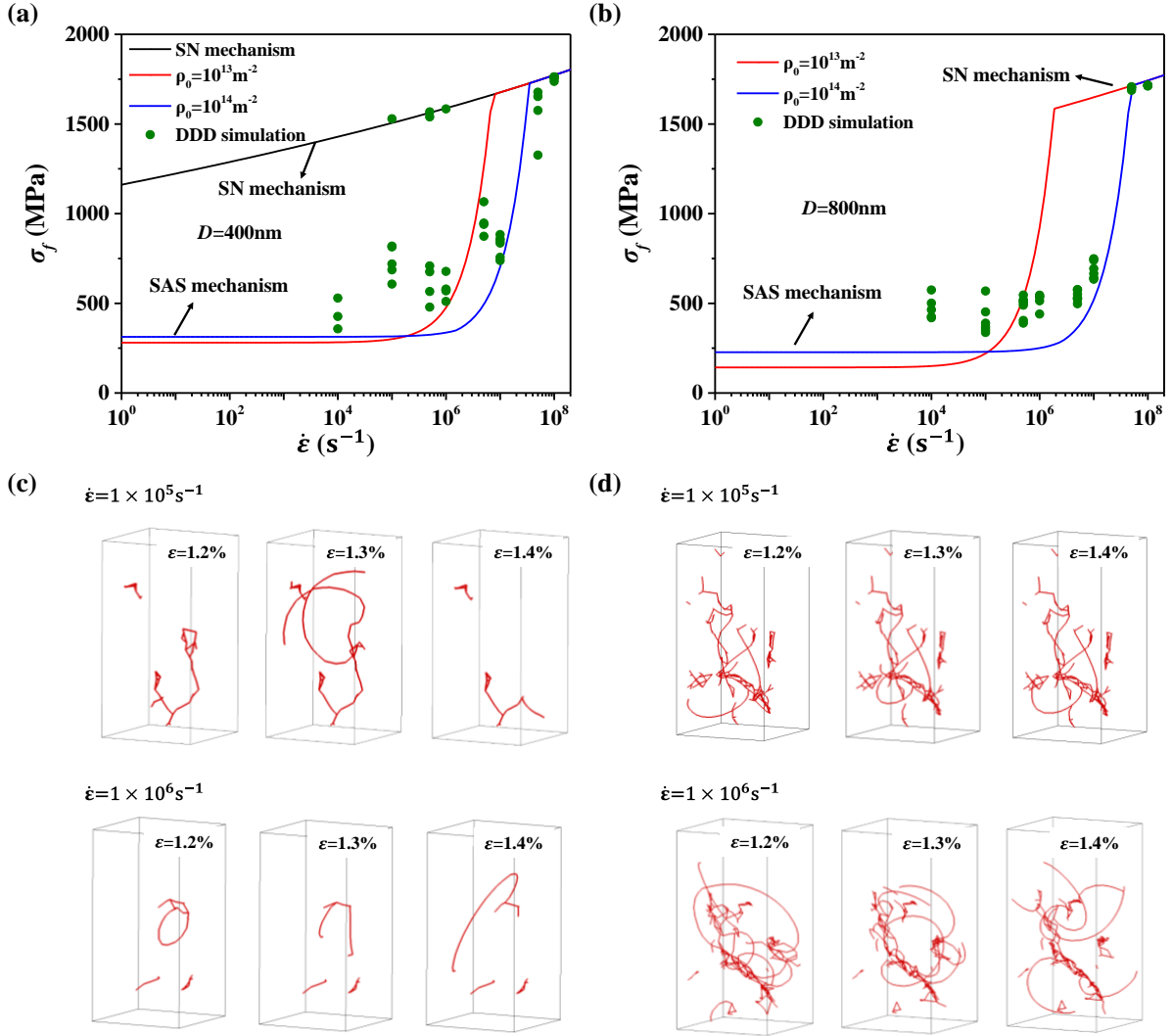

Figure 5 The flow stress of Cu pillars with different initial dislocation densities predicted by the two analytical models (curves) for (a) $D$=400nm and (b) $D$=800nm. The circles are DDD simulation results. The evolution of dislocation structures in (c) $D$=400nm and (d) $D$=800nm.

We further compare our analytical model predictions with the experimental data where the mixed effect of the sample size and the loading rate was tested. This is shown in Figure 6, where the experimental data is taken from [34]. By the evaluation of activation volume, Jennings et al. [34] concluded two distinct plasticity mechanisms, i.e., surface dislocation nucleation and collective dislocation dynamics via SASs. Considering that the dislocation density could only roughly be measured in the experiment as around ~$10^{14}$ m$^{-2}$, we have tested two different sets of dislocation densities (i.e. $5\times10^{13}$ m$^{-2}$ and $1\times10^{14}$ m$^{-2}$) in the analytical models.

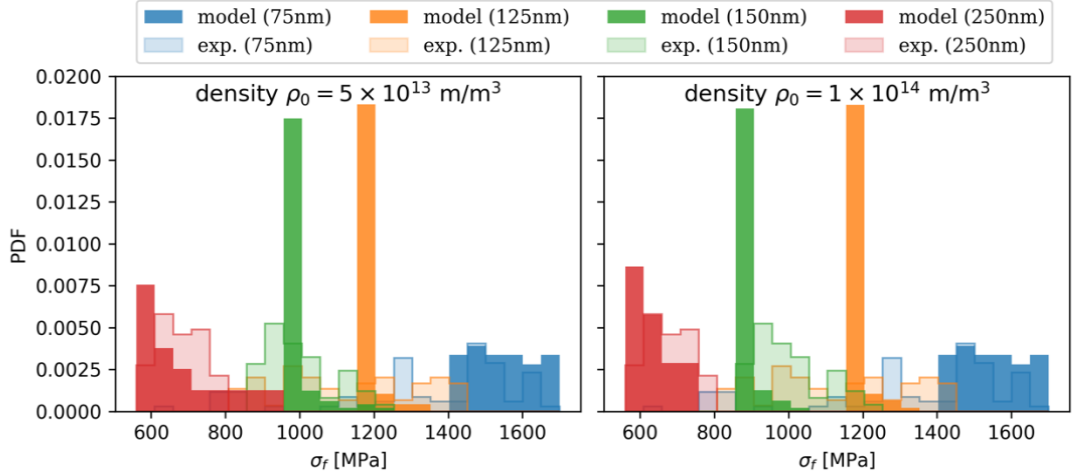

Figure 6 Histogram of the flow stress of Cu pillars comparing the experiment results [34] (bars solid lines) with the model results (filled bars) for two different densities

The histogram shows the flow stresses for different strain rates which is one of the reason for the scatter of the data. We observe, that the simulated flow stresses match the experimental ones for the smallest and largest pillars well. Our model underpredicts only flow stress of the two medium sized pillars. Furthermore, in the simulations the transition from the SAS regime to the SN regime can nicely be seen in the different distribution shapes for the smallest pillar as compared to the larger ones: the SAS have a very well defined stress level under which they operate – resulting in relatively sharp distributions for the pillars with diameter >=125nm. In the smallest pillar, nucleation always happens through surface nucleation, which strongly depends on the external loading rate and the local microstructure near the surface and which therefore results in a much broader distribution. The experimental data is, in this respect, inconclusive which we mainly attribute to the limited data availability.

## 5. Concluding remarks

In the present work, we started by performing DDD simulations to investigate the influence of sample size, strain rate and equilibrium dislocation density on the mechanisms responsible for the evolution of the dislocation microstructure in single crystalline sub-micrometer pillars. Based on this physical insight, two analytical models were established to quantitatively reveal the different effects of surface nucleation and of the nucleation by single armed sources on the plastic deformation behavior. The analytical predictions match well with the DDD simulation results and highlight the effect of sample size, dislocation density (the number of stable SASs) and strain rate. Specifically, for quasi-static loading, the critical size predicted by our analytical models is consistent with the experimental results; finding this point would have required a large number of computationally expensive DDD simulations. Under various strain rates, our analytical model predictions also provide good agreement with DDD simulation results, and shows consistency with the experimental results.

While such a still relatively simple analytical model can clearly not predict spatial details of the dislocation microstructure it draws on the fact that all terms are physically meaningful and almost all parameters are measurable. It is thus useful for both the experimental and the simulation community: For the DDD community the model clearly shows if in the regime surface nucleation happens, which then the respective code must be able to simulate. Alternatively, one could also use the model to design a simulation setting which a priori excludes the SN mechanism, e.g., in order to speed up DDD simulations or because the used code simply is not able to consider surface nucleation. For experimentalists, given material properties and loading conditions, one can get a quick estimate of the sample strength under investigation and perform a simple design-of-experiment study, rather than carrying out computationally expensive DDD simulations.

**Acknowledgments**

This work is supported by the Science Challenge Project (No. TZ2018001) and the National Natural Science Foundation of China (No. 11802310). The work of XL is supported by the National Natural Science Foundation of China (No. 11772334, 11672301), by Youth Innovation Promotion Association CAS (2018022), and by the Strategic Priority Research Program of the Chinese Academy of Sciences (No. XDB22040501). HS and SS acknowledge financial support from the European Research Council through the ERC Grant Agreement No. 759419 MuDiLingo ("A Multiscale Dislocation Language for Data-Driven Materials Science"). The authors are very grateful for the insightful and inspiring discussion with Dr. Yinan Cui.

**Author Contributions**

JH, HS,XL, SS designed the study. JH carried out the 3D DDD simulations with the help and supervision of ZL and ZZ. JH, HS and SS carried out the calculations of analytical models and JH, HS, SS produced all figures. JH, HS and SS wrote the first draft. All authors discussed the results and commented on the manuscript.

**Competing Interests:** The authors declare no competing interests.

**Appendix A．Details of calculation procedure of analytical models**

Under uniaxial loading, the stress in the sample elastically rises until either the SAS or SN gets activated. In the analytical model, based on the loading rate $\dot{\varepsilon}$, the sample size $D$ and equilibrium dislocation density $\rho$, a different mechanism is triggered. The activation of SN process is due to the lack of enough internal dislocation sources (e.g., SAS). Based on the two analytical models proposed in this work, i.e., the SN dominated model and the SAS dominated model, the flow stress can be predicted through the following calculation flow charts, as shown in Figure A1(a) and (b).

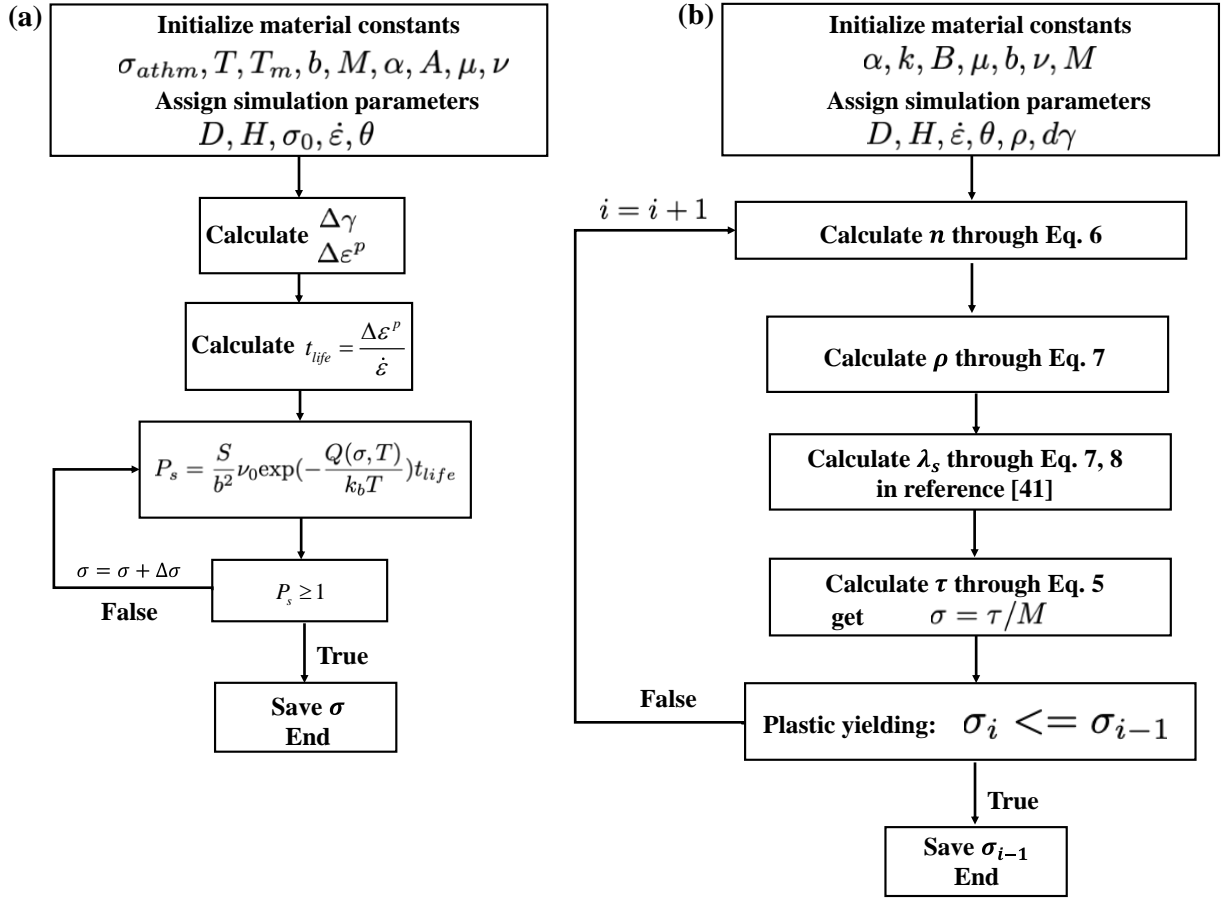

Figure A1 (a) Calculation flow chart of the SN dominated model. (b) Calculation flow chart of the minimum number of SASs and the corresponding flow stress.

Correspondingly, the controlling parameters as well as material properties of copper as listed in Table A1, in which, the parameters associated with the SN analytical model can also be found in [25]. Through these parameters, the flow stress can be obtained when either SN or SAS are activated. The real flow stress/final prediction of the sample is the minimum between the predictions of the two models.

Table A1 Material properties and parameters for copper.

| Material and parameters | Dimension | Value |
|---|---|---|
| Shear modulus ($\mu$) | GPa | 48 |
| Poisson ratio ($v$) |  | 0.34 |
| Burgers vector length ($b$) | nm | 0.256 |
| Mass density ($\rho_m$) | kg/m$^3$ | 8.9×10$^3$ |
| Drag coefficient ($B$) | Pa·s | $2.0\times10^{-5}$ |
| Attempt frequency ($v_0$) | s$^{-1}$ | $1\times10^{13}$ |
| Surface disordering temperature ($T_m$) | K | 700 |
| Modelling temperature ($T$) | K | 300 |
| Athermal stress in SN model ($\sigma_{athm}$) [25] | GPa | 5.2 |
| Proportionality constant for activation energy from ($A$) [25] | eV | 4.8 |
| Fitting exponent for activation energy from ($\alpha_0$) [25] | | 4.1 |
| Dimensionless constant ($k_f$) | | 0.01 |
| Effective mutual annihilation distance ($y_c$) | nm | 0.7 |

**References**

[1] S. Sandfeld, T. Hochrainer, P. Gumbsch, M. Zaiser, Numerical implementation of a 3D continuum theory of dislocation dynamics and application to micro-bending, Philosophical Magazine 90(27-28) (2010) 3697-3728.

[2] S. Sandfeld, T. Hochrainer, M. Zaiser, P. Gumbsch, Continuum modeling of dislocation plasticity: Theory, numerical implementation, and validation by discrete dislocation simulations, Journal of Materials Research 26(05) (2011) 623-632.

[3] Y. Aoyagi, R. Kobayashi, Y. Kaji, K. Shizawa, Modeling and simulation on ultrafine-graining based on multiscale crystal plasticity considering dislocation patterning, International Journal of Plasticity 47 (2013) 13-28.

[4] T. Carvalho Resende, S. Bouvier, F. Abed-Meraim, T. Balan, S.S. Sablin, Dislocation-based model for the prediction of the behavior of b.c.c. materials – Grain size and strain path effects, International Journal of Plasticity 47 (2013) 29-48.

[5] S. Sandfeld, E. Thawinan, C. Wieners, A link between microstructure evolution and macroscopic response in elasto-plasticity: Formulation and numerical approximation of the higher-dimensional continuum dislocation dynamics theory, International Journal of Plasticity 72 (2015) 1-20.

[6] S. Xia, A. El-Azab, Computational modelling of mesoscale dislocation patterning and plastic deformation of single crystals, Modelling and Simulation in Materials Science and Engineering

23(5) (2015) 055009.

[7] A.S. Khan, J. Liu, J.W. Yoon, R. Nambori, Strain rate effect of high purity aluminum single crystals: Experiments and simulations, International Journal of Plasticity 67 (2015) 39-52.

[8] B.L. Hansen, I.J. Beyerlein, C.A. Bronkhorst, E.K. Cerreta, D. Dennis-Koller, A dislocation-based multi-rate single crystal plasticity model, International Journal of Plasticity 44 (2013) 129-146.

[9] Y. Cui, Z. Liu, Z. Zhuang, Quantitative investigations on dislocation based discrete-continuous model of crystal plasticity at submicron scale, International Journal of Plasticity 69 (2015) 54-72.

[10] P. Lin, Z. Liu, Z. Zhuang, Numerical study of the size-dependent deformation morphology in micropillar compressions by a dislocation-based crystal plasticity model, International Journal of Plasticity 87 (2016) 32-47.

[11] A.T. Jennings, M.J. Burek, J.R. Greer, Microstructure versus size: mechanical properties of electroplated single crystalline Cu nanopillars, Physical Review Letters 104(13) (2010) 135503.

[12] M.D. Uchic, D.M. Dimiduk, J.N. Florando, W.D. Nix, Sample dimensions influence strength and crystal plasticity, Science 305 (2004) 986-989.

[13] M.D. Uchic, P.A. Shade, D.M. Dimiduk, Plasticity of Micrometer-Scale Single Crystals in Compression, Annual Review of Materials Research 39(1) (2009) 361-386.

[14] C. Zhou, S.B. Biner, R. LeSar, Discrete dislocation dynamics simulations of plasticity at small scales, Acta Materialia 58(5) (2010) 1565-1577.

[15] Y. Cui, Z. Liu, Z. Zhuang, Theoretical and numerical investigations on confined plasticity in micropillars, Journal of the Mechanics and Physics of Solids 76 (2015) 127-143.

[16] S.-W. Lee, S.M. Han, W.D. Nix, Uniaxial compression of FCC Au nanopillars on an MgO substrate: The effects of prestraining and annealing, Acta Materialia 57(15) (2009) 4404-4415.

[17] S.I. Rao, D.M. Dimiduk, T.A. Parthasarathy, M.D. Uchic, M. Tang, C. Woodward, Athermal mechanisms of size-dependent crystal flow gleaned from three-dimensional discrete dislocation simulations, Acta Materialia 56(13) (2008) 3245-3259.

[18] S.-W. Lee, W.D. Nix, Size dependence of the yield strength of FCC and BCC metallic

micropillars with diameters of a few micrometers, Philosophical Magazine 92(10) (2012) 1238-1260.

[19] Z.W. Shan, R.K. Mishra, S.A. Syed Asif, O.L. Warren, A.M. Minor, Mechanical annealing and source-limited deformation in submicrometre-diameter Ni crystals, Nature materials 7(2) (2008) 115-119.

[20] J.R. Greer, W.D. Nix, Nanoscale gold pillars strengthened through dislocation starvation, Physical Review B 73(24) (2006) 245410.

[21] I. Ryu, W. Cai, W.D. Nix, H. Gao, Stochastic behaviors in plastic deformation of face-centered cubic micropillars governed by surface nucleation and truncated source operation, Acta Materialia 95 (2015) 176-183.

[22] F. Cleri, D. Wolf, S. Yip, S.R. Phillpot, Atomistic simulation of dislocation nucleation and motion from a crack tip, Acta Materialia 45(12) (1997) 4993-5003.

[23] Y. Liu, E. Van der Giessen, A. Needleman, An analysis of dislocation nucleation near a free surface, International Journal of Solids and Structures 44(6) (2007) 1719-1732.

[24] S. Xu, Y.F. Guo, A.H.W. Ngan, A molecular dynamics study on the orientation, size, and dislocation confinement effects on the plastic deformation of Al nanopillars, International Journal of Plasticity 43 (2013) 116-127.

[25] T. Zhu, J. Li, A. Samanta, A. Leach, K. Gall, Temperature and strain-rate dependence of surface dislocation nucleation, Physical Review Letters 100(2) (2008) 025502.

[26] S. Ryu, K. Kang, W. Cai, Predicting the dislocation nucleation rate as a function of temperature and stress, Journal of Materials Research 26(18) (2011) 2335-2354.

[27] M.A. Tschopp, D.E. Spearot, D.L. McDowell, Atomistic simulations of homogeneous dislocation nucleation in single crystal copper, Modelling and Simulation in Materials Science and Engineering 15(7) (2007) 693-709.

[28] T.A. Parthasarathy, S.I. Rao, D.M. Dimiduk, M.D. Uchic, D.R. Trinkle, Contribution to size effect of yield strength from the stochastics of dislocation source lengths in finite samples, Scripta Materialia 56(4) (2007) 313-316.

[29] H. Bei, Y.F. Gao, S. Shim, E.P. George, G.M. Pharr, Strength differences arising from

homogeneous versus heterogeneous dislocation nucleation, Physical Review B 77(6) (2008) 060103.

[30] Y. Guo, Z. Zhuang, X.Y. Li, Z. Chen, An investigation of the combined size and rate effects on the mechanical responses of FCC metals, International Journal of Solids and Structures 44(3-4) (2007) 1180-1195.

[31] J.Y. Zhang, G. Liu, J. Sun, Strain rate effects on the mechanical response in multi- and single-crystalline Cu micropillars: Grain boundary effects, International Journal of Plasticity 50 (2013) 1-17.

[32] S. Sandfeld, M. Zaiser, Pattern formation in a minimal model of continuum dislocation plasticity, Modelling and Simulation in Materials Science and Engineering 23(6) (2015) 065005.

[33] H. Song, D. Dimiduk, S. Papanikolaou, Universality Class of Nanocrystal Plasticity: Localization and Self-Organization in Discrete Dislocation Dynamics, Physical Review Letters 122(17) (2019) 178001.

[34] A.T. Jennings, J. Li, J.R. Greer, Emergence of strain-rate sensitivity in Cu nanopillars: Transition from dislocation multiplication to dislocation nucleation, Acta Materialia 59(14) (2011) 5627-5637.

[35] J. Hu, Z. Liu, E. Van der Giessen, Z. Zhuang, Strain rate effects on the plastic flow in submicron copper pillars: Considering the influence of sample size and dislocation nucleation, Extreme Mechanics Letters 17 (2017) 33-37.

[36] C. Zhou, S. Biner, R. LeSar, Simulations of the effect of surface coatings on plasticity at small scales, Scripta Materialia 63(11) (2010) 1096-1099.

[37] B. Devincre, L.P. Kubin, Mesoscopic simulations of dislocations and plasticity, Materials Science and Engineering 234-236 (1997) 8-14.

[38] P.D. Ispánovity, I. Groma, G. Györgyi, F.F. Csikor, D. Weygand, Submicron plasticity: Yield stress, dislocation avalanches, and velocity distribution, Physical Review Letters 105(8) (2010) 085503.

[39] M.C.Fivel, C.F.Robertson, G.R.Canova, L.Boulanger, Three-Dimensional modeling of indent-induced plastic zone at a mesoscale, Acta Materialia 46(17) (1998) 6183-6194.

[40] G. Yuan, Z. Zhuo, L. Zhanli, Z. Xuechuan, Z. Zhaohui, Characteristic Sizes for Exhaustion-Hardening Mechanism of Compressed Cu Single-Crystal Micropillars, Chinese Physics Letters 27(8) (2010) 086103.

[41] Y. Cui, P. Lin, Z. Liu, Z. Zhuang, Theoretical and numerical investigations of single arm dislocation source controlled plastic flow in FCC micropillars, International Journal of Plasticity 55 (2014) 279-292.

[42] J.P. Hirth, J. Lothe, Theory of dislocations. 2nd ed, New York: Krieger Publishing Company–John Wiley & Sons, Ltd. (1982).

[43] Zhanli Liu, Xiaoming Liu, Zhuo Zhuang, X. You, Atypical three-stage-hardening mechanical behavior of Cu single-crystal micropillars, Scripta Materialia 60(7) (2009) 594-597.

[44] C.R. Weinberger, W. Cai, Computing image stress in an elastic cylinder, Journal of the Mechanics and Physics of Solids 55(10) (2007) 2027-2054.

[45] C.R. Weinberger, W. Cai, Surface-controlled dislocation multiplication in metal micropillars, Proceedings of the National Academy of Sciences of the United States of America 105(38) (2008) 14304-14307.

[46] H. Tang, K. Schwarz, H. Espinosa, Dislocation escape-related size effects in single-crystal micropillars under uniaxial compression, Acta Materialia 55(5) (2007) 1607-1616.

[47] S. Groh, E.B. Marin, M.F. Horstemeyer, H.M. Zbib, Multiscale modeling of the plasticity in an aluminum single crystal, International Journal of Plasticity 25(8) (2009) 1456-1473.

[48] Z.Q. Wang, I.J. Beyerlein, R. Lesar, Dislocation motion in high strain-rate deformation, Philosophical Magazine 87(16) (2007) 2263-2279.

[49] Y. Cui, Z. Liu, Z. Wang, Z. Zhuang, Mechanical annealing under low-amplitude cyclic loading in micropillars, Journal of the Mechanics and Physics of Solids 89 (2016) 1-15.

[50] C. Motz, D. Weygand, J. Senger, P. Gumbsch, Initial dislocation structures in 3-D discrete dislocation dynamics and their influence on microscale plasticity, Acta Materialia 57(6) (2009) 1744-1754.

[51] B. Gurrutxaga-Lerma, D.S. Balint, D. Dini, D.E. Eakins, A.P. Sutton, A dynamic discrete dislocation plasticity method for the simulation of plastic relaxation under shock loading,

Proceedings of the Royal Society A: Mathematical, Physical and Engineering Sciences 469(2156) (2013) 1-24.

[52] Y. Cui, G. Po, Y.-P. Pellegrini, M. Lazar, N. Ghoniem, Computational 3-dimensional dislocation elastodynamics, Journal of the Mechanics and Physics of Solids 126 (2019) 20-51.

[53] W.D. Nix, J.R. Greer, G. Feng, E.T. Lilleodden, Deformation at the nanometer and micrometer length scales: Effects of strain gradients and dislocation starvation, Thin Solid Films 515(6) (2007) 3152-3157.

[54] S.H. Oh, M. Legros, D. Kiener, G. Dehm, In situ observation of dislocation nucleation and escape in a submicrometre aluminium single crystal, Nature materials 8(2) (2009) 95-100.

[55] Z. Wang, Q. Li, Z. Shan, J. Li, J. Sun, E. Ma, Sample size effects on the large strain bursts in submicron aluminum pillars, Applied Physics Letters 100(7) (2012) 071906.

[56] C. Zhou, I.J. Beyerlein, R. LeSar, Plastic deformation mechanisms of FCC single crystals at small scales, Acta Materialia 59(20) (2011) 7673-7682.

[57] J.A. El-Awady, Unravelling the physics of size-dependent dislocation-mediated plasticity, Nature communications 6(5926) (2015) 1-9.

[58] A.T. Jennings, C. Gross, F. Greer, Z.H. Aitken, S.W. Lee, C.R. Weinberger, J.R. Greer, Higher compressive strengths and the Bauschinger effect in conformally passivated copper nanopillars, Acta Materialia 60(8) (2012) 3444-3455.

[59] D. Kiener, A.M. Minor, Source-controlled yield and hardening of Cu(100) studied by in situ transmission electron microscopy, Acta Materialia 59(4) (2011) 1328-1337.

[60] D. Kiener, A.M. Minor, Source truncation and exhaustion: insights from quantitative in situ TEM tensile testing, Nano Letters 11(9) (2011) 3816-3820.

[61] Z. Shan, In Situ TEM Investigation of the Mechanical Behavior of Micronanoscaled Metal Pillars, JOM 64(10) (2012) 1229-1234.